\documentstyle[twocolumn,aps,prb,epsfig]{revtex}
\begin{document}
\draft
\twocolumn[\hsize\textwidth\columnwidth\hsize\csname @twocolumnfalse\endcsname%
%
%

\title{Thermodynamics of Random Ferromagnetic Antiferromagnetic 
Spin-1/2 Chains}

\author{Beat Frischmuth$^1$, Manfred Sigrist$^2$, Beat Ammon$^1$ 
and Matthias Troyer$^3$}

\address{$^1$ Institute of Theoretical Physics, ETH Hoenggerberg, 
CH-8093 Zuerich, Switzerland \\
$^2$ Yukawa Institute for Theoretical Physics, Kyoto University, 
Kyoto 606-01, Japan \\
$^3$ Institute for Solid State Physics, University of Tokyo, 
Roppongi 7-22-1, Tokyo 106, Japan 
}

\date{\today}
\maketitle

\begin{abstract}
  Using the quantum Monte Carlo Loop algorithm, we calculate the
  temperature dependence of the uniform susceptibility, the specific
  heat, the correlation length, the generalized staggered
  susceptibility and magnetization of a spin-1/2 chain with random
  antiferromagnetic and ferromagnetic couplings, down to very low
  temperatures. Our data show a consistent scaling behavior in all the
  quantities and support strongly the conjecture drawn from the
  approximate real-space renormalization group treatment.  A
  statistical analysis scheme is developed which will be useful for
  the search of scaling behavior in numerical and experimental data of
  random spin chains.
\end{abstract}
\vskip2pc] \narrowtext

\section{Introduction}

Over many decades one-dimensional (1D) quantum spin systems have
attracted much interest. Despite of their simplicity they show a
wealth of physical properties which provide many keys to the
understanding of phenomena appearing also in other strongly correlated
systems.  Since the discovery of various quasi-1D materials the study
of 1D spin systems, mainly based on the Heisenberg model and its
variations, is also of experimental relevance. Examples of such
materials include the so-called NINO
(Ni(C$_3$H$_{10}$N$_2$)$_2$NO$_2$(Cl)$_4$)), NENP
(Ni(C$_2$H$_8$N$_2$)$_2$NO$_2$(Cl)$_4$)),\cite{renard,miller} and
Sr$_3$CuPtO$_6$.\cite{wilkinson} In particular, the latter system
belongs to a class of compounds which is compositionally very
flexible.  It is possible to produce compounds of the form
Sr$_3$MNO$_6$ in various combinations with M$=$Cu, Mg, Zn, Yb, Na, Ca,
Co, and N$=$Pt, Ir, Rh, Bi. Other examples are the spin ladder
systems, such as Sr$_{n-1}$Cu$_{n+1}$O$_{2n}$ ($n+1$, the number of
ladder legs) or CaV$_2$O$_5$.\cite{dagotto}

Disorder effects play a particularly important role in quasi-1D systems,
because even small deviations from regularity often destabilize the
pure phases.\cite{doty} Real experimental systems naturally contain
impurities and other types of disorder. Therefore it is very important
to understand the influence of disorder on the properties of such
systems in order to interpret experimental results. 

This is one reason why in recent years random spin systems have been
investigated intensively. But also from the theoretical point of view,
random spin systems are interesting since they provide a simple model
to study the interplay between quantum effects and disorder.  There
are various realizations of quasi-one-dimensional random spin systems
in nature. One class is represented by compounds like
Sr$_3$CuPt$_{1-x}$Ir$_x$O$_6$.\cite{nguyen} While the pure compounds
Sr$_3$CuPtO$_6$ ($x=0$) and Sr$_3$CuIrO$_6$ ($x=1$) are
antiferromagnetic (AFM) and ferromagnetic (FM), respectively, the alloy
Sr$_3$CuPt$_{1-x}$Ir$_x$O$_6$ contains both AFM and FM couplings whose
fraction is simply related to the concentration $x$ of Ir. A
corresponding minimal model of this type is the spin-1/2 Heisenberg
chain, where the nearest neighbor exchange coupling is $+J$ or $-J$
with certain probabilities.\cite{furusaki} Another related example is
realized in the low-temperature regime of randomly depleted AFM
spin-1/2 Heisenberg ladders. Two-leg Heisenberg ladders have a
resonating valence bond ground state which has short-range singlet
correlation and a spin excitation gap.\cite{dagotto} If spins are
depleted at random the low-energy properties are changed drastically.
Each site which lost a spin is accompanied by an effective spin 1/2,
and the residual interaction among these spins is randomly FM or AFM
with a wide distribution of coupling strengths.\cite{fukuyama,sigrist}
Recently, it was also shown that random AFM
spin-1 chains including next-nearest neighbor couplings can generate
effective ferromagnetic couplings in the low-energy regime so that
they should behave similarly at low-temperature.\cite{yang}

In this work we focus on random FM-AFM spin chains with the full spin
rotation symmetry.  One of the most insightful techniques to study
such systems is the real space renormalization group (RSRG)
method.\cite{dasgupta,ma,hirsch1,hirsch2,fisher,west1,west} This
method was introduced to study random AFM spin-$1/2$
chains,\cite{dasgupta,ma,hirsch1,hirsch2,fisher} and was recently
adapted to the study of the class of systems introduced above.
\cite{west1,west} The basic idea of the RSRG method is the iterative
decimation of degrees of freedom by successively integrating out the
strongest bonds in the spin chain. In this way the distribution of
coupling strengths and spin sizes is renormalized. For many cases an
universal fixed point distribution is reached in the low-energy limit.
The random AFM spin-$1/2$ chain belongs to a universality class
different from that of the random FM-AFM spin chain. The former is
characterized by a {\it random singlet phase} where each spin tends to
form a singlet with one other spin even over very large distances. For
the FM-AFM spin chain, however, spins correlate to form effective
spins whose average size grows with lowering of the energy
scale.\cite{west1,west,hida} A result of the RSRG method is that the
average number of original spins included in a single effective spin
for a given (energy or) temperature $ T $ scales as $ \bar{l} \propto
T^{- 2 \alpha} $, and the average spin size $ \bar{S} \propto
T^{-\alpha} $ for $ T \to 0$.\cite{west1,west}

We start our analysis with a brief review of the RSRG scheme and its
results\cite{west1,west} for the random FM-AFM spin chain which shows
a universal scaling behavior at low-temperatures (section 4.2). The
RSRG treatment, however, relies on certain approximations which have
not so far been independently tested. We use, therefore, a different
approach to examine the random FM-AFM spin-1/2 chain and the validity
of the scaling assumption underlying the RSRG scheme.\cite{ich3} For
the analysis of the numerical data obtained by the continuous time
version of the quantum Monte Carlo (QMC) loop
algorithm\cite{evertz,beard} we introduce a statistical description of
the low-temperature properties (section 4.3).  Various quantities such
as the uniform susceptibility, specific heat, generalized staggered
susceptibility, correlation and generalized staggered magnetization
are discussed in section 4.4. We find excellent agreement between our
analysis and the RSRG treatment by Westerberg {\it et
  al.}.\cite{west1,west} Beyond the consistent evaluation of the
scaling exponent $ \alpha $, our simulation gives quantities mentioned
above over a wide range of temperatures and our discussion provides a
technique to analyze (numerical or experimental) data for the
low-temperature scaling regime.

\section{Real space renormalization group treatment}

The real space renormalization group scheme used by Westerberg {\it et
  al.}\cite{west1,west} can be applied to the large class of random
spin systems with a Hamiltonian of the form
\begin{equation}\label{HamRG}
H=\sum_i J_i\vec{S}_i\cdot\vec{S}_{i+1},
\end{equation}
where the strength and sign of $J_i$ as well as the size of the spins
$S_i$ are randomly distributed according to the distributions $P(J)$
and $\tilde P(S)$, respectively. The basic idea of the RSRG method is
the iterative decimation of degrees of freedom by integrating out
successively the strongest bonds in the spin chain. The RG step will
now be described in more detail.

For this purpose a bond in the chain is defined as two neighboring
spins and the coupling $J$ connecting them. If a bond is isolated from
the rest of the chain, it would form a local ground state of maximum
($J<0$) or minimum ($J>0$) spin with an energy gap $\Delta$ to the
first excited multiplet. For a FM bond $\Delta=-J(S_L+S_R)$, while for
an AFM bond $\Delta=J(|S_L-S_R|+1)$, where $S_L$ and $S_R$ are the
left and right spins of the bond, respectively. Now we focus on the
strongest bond in the chain, defined as the bond with the largest gap
$\Delta=\Delta_0$.  If the distribution of gaps is broad (this is the
case if the coupling strength and/or the spin-size distribution is
broad), the gaps of the two neighboring bonds, $\Delta_1$ and
$\Delta_2$, are typically much smaller than $\Delta_0$ so that the two
spins $S_L$ and $S_R$, to a good approximation, lock into their local
ground state.  Consequently, the bond ${\Delta_0,S_L,S_R}$ is replaced
by a single effective spin $S'=|S_L \pm S_R|$ representing the local
ground state of minimum (AFM) or maximum (FM) spin. The weaker
neighboring bonds are then taken into account perturbatively in
$\Delta_{1,2}/\Delta_0$, leading to an effective interaction between
the spins $S_1$, $S'$, and $S_2$ (see Fig.~\ref{RGstep}). The spin
rotation symmetry is preserved in this procedure, and to first order
in $\Delta_{1,2}/\Delta_0$ no next-nearest neighbor interactions are
generated.  At this point, however, it is important to note that the
crucial assumption of $\Delta_1,\ \Delta_2 \ll \Delta_0$ is probable,
but uncontrolled and independent tests are necessary to examine
whether this assumption and, hence, the RSRG results are correct
(section 4.3 and 4.4).

Integrating out the strongest bond successively in the manner
described above preserves the the form of the Hamiltonian
Eq.~(\ref{HamRG}) but changes the coupling strength distribution, the
spin distribution and in particular reduces the energy scale
$\Delta_0$.  We expect that after many iteratively performed
decimation steps the distributions flow to a fixed-point distributions
with universal scaling behavior. According to Westerberg {\it et
  al.},\cite{west1,west} these fixed point distributions are universal
for a large class of initial $P(J)$ and $\tilde P(S)$, namely if the
initial gap distribution $\hat P(\Delta)$ is regular or less singular
than $\hat P(\Delta)\sim \Delta^{-0.7}$.

\vspace*{3mm}
\begin{figure}
\begin{center}
\epsfxsize=45mm
\epsffile{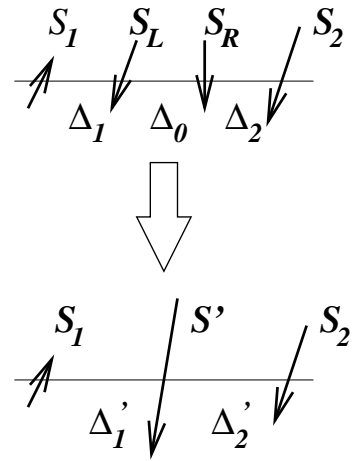}
\vspace*{3mm}
\caption[]{Schematic picture of one step in the RSRG treatment of 
Westerberg {\it et al.}.\cite{west1,west}}\label{RGstep}
\end{center}
\end{figure}

The RSRG results\cite{west1,west} show that for this whole class of
distributions the average number $\bar{l}$ of original spins included
in one effective large spin diverges with $\Delta_0^{-2 \alpha}$ as
$\Delta_0\rightarrow 0$ with an universal exponent $\alpha=0.22\pm
0.01$. The total spin of an effective large spin is the sum of the
participating original spins, where each spin enters the sum with the
(opposite) sign as its neighbor if the coupling is ferromagnetic
(antiferromagnetic). By a random walk argument one finds for the
average total quantum spin number $\bar{S}$
\begin{equation}
\bar{S}\propto\sqrt{\bar{l}}\propto\Delta_0^{-\alpha}.
\end{equation}

\begin{figure}[h]
\twocolumn[\hsize\textwidth\columnwidth\hsize\csname @twocolumnfalse\endcsname%
{\begin{center}
\epsfxsize=130mm
\epsffile{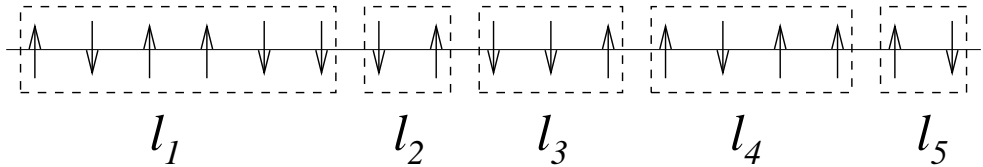}
\end{center}
\caption[]{Possible cluster forming at given temperature $T$. The 
cluster lengths $l_1,\ l_2,\ \ldots$ are exponentially distributed 
with an average value $\bar{l}=\lambda^2 (T/J_0)^{-2 \alpha}$}
\label{cluster}}
\vskip2pc]
\end{figure}
\noindent{At finite temperature $T$ the RG flows will stop at 
  $\Delta_0\sim k_B T$.  All bonds with a gap larger than $\Delta_0$
  are essentially frozen to form effective large spins which are, in
  turn, more or less independent from each other. These free effective
  spins lead to a Curie like behavior of the magnetic uniform
  susceptibility $\chi$ per unit length for $T\rightarrow 0$, since}
the scaling behavior of $ \bar{S}^2 $ and $ \bar{l} $ is identical,
\begin{equation}\label{chi_RSRG}
\frac{\chi}{L}=\frac{\mu^2}{3 k_B T} \frac{\bar{S}^2}{\bar{l}} \propto
\frac{1}{T}\frac{\Delta_0^{-2 \alpha}}{\Delta_0^{-2 \alpha}} = \frac{1}{T}.
\end{equation}
For the entropy $ \sigma $ (per original spin) we can argue that only
the free effective spins contribute such that 

\begin{equation}\label{sig_RSRG}
\sigma = \frac{\bar{l}} k_B {\rm ln}(2 \bar{S} +1) \propto T^{2\alpha}
{\rm ln} T 
\end{equation}
in leading order, which is identical to that of the specific heat $ C
$ per spin.

In summary we have the following physical picture of the RSRG scheme.
At a given finite temperature $T$, the original spins are grouped in 
completely correlated clusters forming an effective spin. These
clusters are independent and have an average length $\bar{l}\propto
T^{-2\alpha}$. This will be the basis of our statistical
approach for the discussion of the thermodynamic properties.

\section{Methods}
\subsection{Quantum Monte Carlo calculations}

For the numerical simulation we use as model a
FM-AFM spin-$1/2$ chain
\begin{equation}
H=\sum_{i}J_i\vec{S}_i \vec{S}_{i+1},
\end{equation}
where $\vec{S}_i$ is the $i^{th}$ spin of the chain and $J_i$, the
coupling strength of the $i^{th}$ bond, takes random values of both
positive and negative sign. We assume
the bond distribution 
\begin{equation}\label{dist} 
P(J)=\left\{ \begin{array}{ll}
                \frac{1}{2 J_0} & \mbox{$-J_0< J < J_0$} \\
                0               & \mbox{\rm otherwise,}
             \end{array}
     \right.
\end{equation}
where $J_0$ is the maximal coupling setting the energy scale.

According to Westerberg {\it et al.},\cite{west} the low energy
behavior of the random FM-AFM chain is independent of the initial
distribution as long as $J^{y_c} P(J)$ is regular for $J\rightarrow 0$
with $y_c \approx 0.7$. Our choice particular distribution [Eq.
(\ref{dist})] is guided by the feasibility for the numerical analysis
of the scaling regime. This does not restrict the validity of the
conclusions since the scaling behavior should be universal for all
distributions in above class.  Distribution functions closer to those
of the real systems mentioned in the introduction, on the other hand,
while interesting to study in connection with experiments on real
material \cite{ammon}, are not suitable for an accurate analysis of
the scaling regime, since the scaling regime occurs at much lower
temperature relative to the characteristic coupling strength $J_0$.

Using the QMC loop method\cite{evertz,beard}, we simulate 100
samples of spin chains of 400 sites in a temperature range down to
temperatures as low as $T=J_0/1000$. In the QMC loop algorithm the
calculations are performed directly with a Trotter time interval
$\Delta\tau=0$, so no extrapolation in the Trotter time interval is
necessary (see section 2.7).  To obtain good statistics, we consider
100 different realizations of random coupling spin chains,
 following the distribution in
Eq. (\ref{dist}).  The temperature dependences of the physical
properties are calculated for each chain separately, and then averaged
over the 100 samples.

\subsection{Statistical cluster analysis}

The fact that the correlation of spins occurs within definite clusters
permits the use of a special statistical approach for the discussion
of the thermodynamic properties. Original spin-1/2s belonging to the
same cluster are completely correlated, while correlations between
original spins belonging to different cluster can assumed to be zero
(c.f.  Fig.~\ref{cluster}). Therefore, the contributions of these
frozen clusters to a physical observable $\mathcal{A}$ can be
considered independently and be estimated simply as the zero-temperature
(ground state) value within each cluster (depending only on their length
$l$).

Performing a statistical analysis of the cluster length $l$ allows to
go beyond the leading order in the temperature dependence of $\chi$
and $\sigma$, as predicted by the RSRG [Eq. (\ref{chi_RSRG}) and Eq.
(\ref{sig_RSRG})], as well as to calculate also the $T$-dependence of
other physical quantities. Namely, the low temperature dependence of
physical observable $\mathcal{A}$ per site in a random AFM-FM spin-1/2
chain of length $L\gg1$ can be expressed as
\begin{equation}
A(T)=\frac{1}{L}\sum_{l=1}^\infty n_T(l)\;l\mbox{$\mathcal{A}$}_l,
\end{equation}
where $n_T(l)$ is the number of clusters of length $l$ and
$\mbox{$\mathcal{A}$}_l$ is the zero-temperature (ground state) value
of $\mathcal{A}$ per site for such a cluster. As $L$ goes to infinity,
we define the probability distribution $p_T(l)$ of the cluster length
$l$ to appear at given temperature $T$, as $p_T(l)\equiv n_T(l)/\sum_l
n_T(l)$. The total number of clusters $\sum_l n_T(l)$ is equal to
$L/\bar{l}$, where $\bar{l}$ is the average cluster length. Hence
\begin{equation}
A(T)=\frac{1}{\bar{l}}\sum_{l=1}^\infty p_T(l)\;l\mbox{$\mathcal{A}$}_l.
\end{equation}
As mentioned above, the RSRG approach predicts
a powerlaw scaling behavior for the temperature dependence of
$\bar{l}$,
\begin{equation}\label{length}
\bar{l}=\lambda^2\left(\frac{T}{J_0}\right)^{-2 \alpha},
\end{equation}
where $\lambda$ is a dimensionless proportionality factor.

Let us now consider the distribution of cluster lengths $ l $ for a
given temperature. It is natural to assume that the bonds freeze
uncorrelated to each other, i.e. each bond is frozen with a certain
probability independent of the location within the chain. Thus, the
distribution  $ p_T(l) $ of the cluster length $ l $ has an
exponential form.  Its average value $\bar{l}$ is given by
Eq. (\ref{length}) and, hence, we will write $p_{\bar{l}}(l)\equiv p_T(l)$
in the following. The distribution $p_{\bar{l}}(l)$ has to be
normalized for a sum over $ l $ from 1 to $ \infty$ and reads in its
discrete form
\begin{equation}\label{distpdisc}
p_{\bar{l}}(l)=(e^\kappa-1) e^{-\kappa l}, \ \ \mbox{\rm with} \
\kappa=\log (1+\frac{1}{\bar{l}-1}). 
\end{equation}

The distribution $\rho_l(S)$ of the spin size of a cluster of fixed
length $l$ also plays an important role, since the contribution
$\mbox{$\mathcal{A}$}_l$ to many physical observables can be
determined using $\rho_l(S)$. The effective spin of a cluster of
length $ l $ is equal to the ground state spin quantum number $S$. For
spin-1/2 degrees of freedom coupled randomly by FM or AFM bonds the
probability distribution for these quantum numbers is given by
(an explanation is presented below)
\begin{equation}\label{distSdisc}
\rho_{l}(S) = \frac{l!}{(l/2+S)! (l/2-S)!}
                \frac{1}{2^l} \\ (2 - \delta_{S,0}),
\end{equation}
where for even $l$, $ S $ is integer with $0\leq S \leq l /2 $ and for
odd $l$, $S$ is half-integer with $\frac{1}{2}\leq S \leq l/2$.
Because the system contains no frustrating couplings, the spin quantum
number $ S $ is identical to the total spin of the corresponding
completely correlated cluster of classical spins.\cite{marshall} The
total spin of a classical cluster of length $l$, however, can easily
be calculated. It is the sum of all original spins 1/2, where each
spin enters the sum with the same (opposite) sign as its neighbor if
the bond is ferromagnetic (antiferromagnetic).  The summation
therefore can be visualized by a random walk on a 1D lattice. A
ferromagnetic (antiferromagnetic) bond corresponds to a step of one
lattice constant to the left (right). The random walk contains of $l$
steps and the distance between the end point of the random walk and
the origin in the 1D lattice is equal to $2S$ and the distribution
$\rho_l(S)$ [Eq.  (\ref{distSdisc})] gives the probability for such a
random walk.

The two distributions $p_{\bar{l}}(l)$ [Eq. (\ref{distpdisc})]
and $\rho_l(S)$ [Eq. (\ref{distSdisc})] together with the scaling
behavior of $\bar{l}$ [Eq. (\ref{length})] are sufficient to calculate 
the low-temperature thermodynamics of various physical quantities.
This will be done in the next section for the uniform
susceptibility, specific heat, entropy and correlation length. 

\section{Results}

\subsection{Intermediate temperature regime}

Before turning to the main point of this chapter, the study of the
universal low-$T$ scaling regime, we will concentrate on the
intermediate temperature range of the random FM-AFM spin-1/2 chains.
The behavior in this range and the way entering the universal low-$T$
are both strongly dependent on the initial coupling distribution
function $P(J)$. To study these points, in this subsection, we
choose different initial coupling distribution functions $P(J)$ having
a width $\int_{-\infty}^\infty dJ |J| P(J) = J_0/2$ and being
normalized to one ($\int_{-\infty}^\infty dJ P(J)=1$).

First we investigate the uniform Curie constant $T\chi(T)$, as a
measure for the still uncorrelated spins at a given temperature,
considering random FM-AFM spin chains of four sites with open boundary
conditions.  For these systems the bond average can be performed
exactly and the calculations take only little time. Thus a more
extended study of the dependence of the behavior on the initial
coupling distribution $P(J)$ in the intermediate temperature range is
possible.  It is clear that finite size effects allow only a
qualitative study of this problem, but the main characteristic
features can already be seen in the 4-site chain.

\begin{figure}[t]
\epsfxsize=85mm
\begin{center}
\epsffile{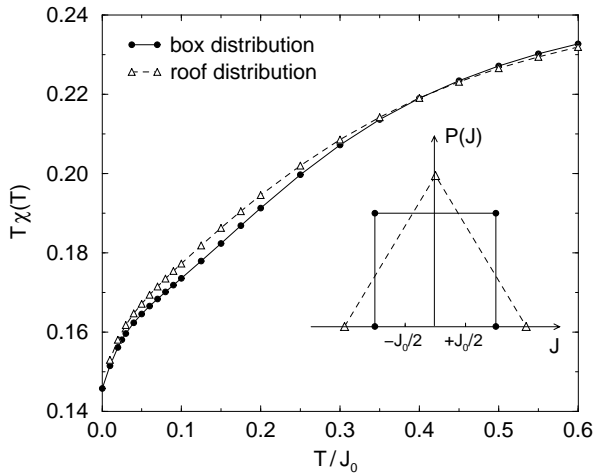}
\end{center}
\caption[]{Temperature dependence of the uniform Curie constant $T\chi(T)$ 
for 4-site random spin chains with initial coupling distribution function 
$P(J)$ shown in the inset (box and roof form).}\label{curie_roof}
\end{figure}

The principal process leading to the dependence of the uniform Curie
constant on the initial coupling distribution function is the
following: For a given temperature the bonds which had a coupling $ J
\gtrsim T$ are frozen either into a singlet (AFM bond) or a triplet (FM
bond) configuration.  Thus, an uncorrelated bond contributes 1/2 to $ T
\chi $, a triplet bond 2/3 and a singlet zero (we set the Bohr
magneton to unity). Thus, the average contribution per bond is reduced
if the FM and AFM couplings occur with the same probability, as
considered here.  The decrease of $ T \chi $ corresponds to the rate
of freezing of bonds as the temperature is lowered, i.e. the constant
distribution function leads to a linear temperature dependence, as
observed for the box distribution (see Fig.~\ref{curie_roof} and
Fig.~\ref{curie_box}).  For the roof distribution (shown in the inset
of Fig.~\ref{curie_roof}), compared with the box distribution, the
deviation from the free spin-1/2 limit of $1/4$ starts at higher
temperature, since the largest possible coupling in the chain is
larger for the roof than for the box distribution. In the intermediate
temperature range ($0.1 J_0 \lesssim T \lesssim 0.5 J_0$) the
decrease of the Curie constant for the roof distribution with
decreasing $T$ is first smaller than for the box distribution, but
then increases and entering the low temperature regime it is again
larger than for the box distribution, as expected from the shape of
both initial coupling distributions (see Fig.~\ref{curie_roof}).
Here, the low-$T$ regime means the temperature range, where not only
single bonds are frozen, but all four spins start to be correlated and
renormalization effects of the distribution function becomes visible.

\begin{figure}[h]
\epsfxsize=85mm
\begin{center}
\epsffile{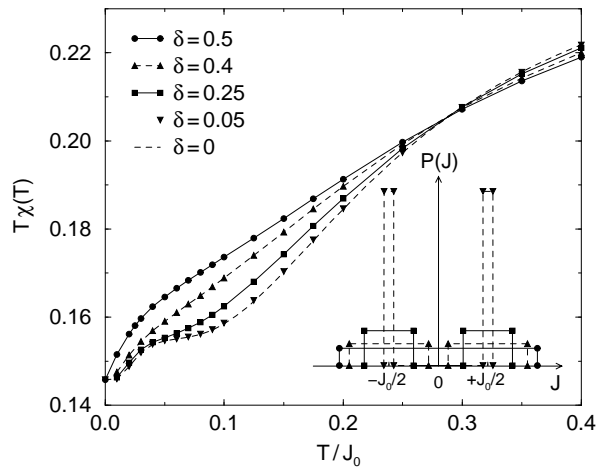}
\end{center}
\caption[]{Temperature dependence of the uniform Curie constant $T\chi(T)$ 
  for 4-site random spin chains with initial coupling distribution
  function Eq. (\ref{dist1}) shown in the inset. The difference
  between the Curie constant for the case $\delta=0.05$ and $\delta=0$
  are smaller then the symbols. }\label{curie_peak}
\end{figure}

In a second step we analyze the effect of an initial coupling
distribution $P(J)$ which is no longer broad but rather peaked at some
values $J\neq0$. In Fig.~\ref{curie_peak} the temperature dependence of
the Curie constants for 4-site chains with initial coupling
distributions
\begin{equation}\label{dist1} 
P(J)=\left\{ \begin{array}{ll}
                \frac{1}{4 \delta J_0} & \mbox{\ if $(\frac{1}{2}-\delta) J_0
                  < |J| < (\frac{1}{2}+\delta) J_0$} \\
                0               & \mbox{\ \rm otherwise,}
             \end{array}
     \right.
\end{equation}
for different $\delta=0, 0.05, 0.25, 0.4, 0.5$ are shown. The case
$\delta=0$ corresponds to a random chain with couplings $\pm J_0/2$,
while the case with $\delta=0.5$ represents the box distribution Eq.
(\ref{dist}).  The latter was already discussed above. The other broad
distribution $\delta=0.4$ leads to a very similar behavior in $T\chi$
to the box distribution and the small differences can be explained by
the same argument as before. The random spin chains with peaked
initial $P(J)$ ($\delta=0.05 \mbox{ and } 0$), however, behave
completely different in the intermediate temperature range. As all
couplings are approximately of the same strength, the picture of
successively freezing of the strongest bond with lowering
temperature no longer applies. According to Furusaki 
{\it et al.}\cite{furusaki}
in such random chains, in the intermediate temperature
regime all sequences of only AFM and of only FM bonds lock together,
forming new effective spins of random spin size and which are coupled
very weakly according to a new and broad coupling distribution. This
forming of weakly interacting effective spins is indicated by the
plateau in the uniform Curie constant (see Fig.~\ref{curie_peak}). As
the temperature is lowered further, the effective spins start to
correlate and the Curie constant decreases again.

The study of the 4-site chains, giving an idea of how the universal
scaling regime is reached for different $P(J)$, clearly shows the
advantages of our choice of a broad initial distribution $P(J)$ [Eq.
(\ref{dist})] for the numerical QMC study of the low-temperature
scaling regime.  Reducing the energy scale, the broad distributions
are renormalized quite rapidly to the fixed point coupling
distribution, while in the case of the peaked distributions, first new
effective spins have to be formed before entering the universal
low-temperature scaling regime.

Whereas the thermodynamic properties in the intermediate temperature range
and the crossover temperature to the scaling regime depend strongly on
the initial coupling distribution, the low-temperature regime of
random FM-AFM spin-1/2 chains itself should show universal behavior.
This will be analyzed in the following, based on the scaling
assumption of the RSRG method, using the statistical cluster analysis,
introduced in section 4.3, as well as the QMC results for the box
distribution Eq. (\ref{dist}).

\subsection{Low-temperature Curie behavior}

\begin{figure}[h]
\epsfxsize=85mm
\begin{center}
\epsffile{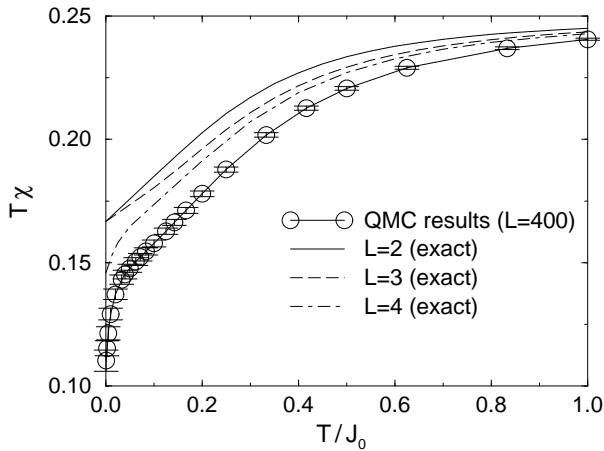}
\end{center}
\caption[]{QMC results for the temperature dependence of the uniform 
  Curie constant of random FM-AFM spin-1/2 chains of length $L=400$
  with a bond distribution according to (\ref{dist}). Also shown are
  the results for $L=2,\,3$ and 4 calculated exactly by integrating
  over the distribution for each bond. For the small clusters open
  boundary conditions are used to avoid strong frustrations.}
  \label{curie_box}
\end{figure}

An important result of the RSRG approach is that the uniform
susceptibility approaches a Curie-like behavior in the limit $ T \to
0 $. It is possible to determine the limiting Curie constant from the
initial spin size distribution and the ratio of FM to AFM 
bonds\cite{west1,west}. The low-temperature deviations from this Curie
constant give important information on the scaling properties
mentioned above. For the scaling analysis we calculate the product $ T
\chi(T) $ with the QMC simulation over a temperature range, $ J_0/1000
\leq T \leq J_0 $ as shown in Fig.~\ref{curie_box}. According to the RSRG
scheme, with lowering temperature gradually the spins correlate within
clusters of growing length $ l $ (see Fig.~\ref{cluster}). Each such
cluster represents an effective spin.  In our case this leads to an
effective diminishing of the spin degrees of freedom such that $ T
\chi(T) $, as a measure for the still uncorrelated spins at a given
temperature, decreases monotonically from the large-$T$ value of 1/4
(the free spin-1/2 limit). The linear dependence in the intermediate
temperature range reflects the initial bond distribution $ P(J) $ in
Eq. (\ref{dist}), as discussed above for the 4-site chains.


We now turn to the analysis of the low-temperature behavior of $ \chi
$. Since each cluster behaves independently we can calculate its
contribution to the susceptibility and average over all cluster sizes 
and spin configurations. Thus for clusters of the average length $
\bar{l} $ the Curie ``constant'' is given by

\begin{equation}\label{curie0}
T\chi(T)=\frac{1}{\bar{l}} \sum_{l=1}^{\infty} p_{\bar{l}}(l)
\;\mbox{$\mathcal{C}$}_l, \end{equation}
where $\mbox{$\mathcal{C}$}_l$ is the average value of Curie constant
per site for a finite, completely correlated cluster of length $l$
which is given by the average over all possible spin quantum numbers
appearing in such a cluster
\begin{equation}\label{curie1}
\mbox{$\mathcal{C}$}_l = \frac{1}{3l} \sum_{S=0}^{l/2} \rho_{l}(S)
S(S+1).
\end{equation}
If $ \bar{l} \gg 1 $, it is justified that we use
the distribution functions $ \rho_l (S) $ and $ p_{\bar{l}}(l) $ in
their continuum approximations 
\begin{equation} \begin{array}{l} \label{dist_cont}
\rho_l (S) = 2 \sqrt{\frac{2}{\pi l}} e^{-2S^2/l} \\ \\ 
p_{\bar{l}}(l)= \frac{e^{1/(\bar{l} -1)}}{\bar{l} -1} e^{-l/(\bar{l}
  -1)} \\
\end{array} 
\end{equation}
for $ l \geq 1 $, as approximation of Eq. (\ref{distpdisc}) and Eq.
(\ref{distSdisc}) -- a comparison, done in Appendix A,  of the following
results with numerical results obtained using the discrete distribution
shows only minor deviations
even for small $ \bar{l}$. These two distributions
[Eq. (\ref{dist_cont})] are now used to calculate the low-temperature
behavior of $T\chi$

\begin{eqnarray} \label{curie2}
T \chi(T) & = & \frac{1}{\bar{l}} \int_1^{\infty} dl\,p_{\bar{l}} (l) l 
\mbox{$\mathcal{C}$}_l \nonumber \\
        &= &\frac{1}{12} + \frac{\bar{l}^{-1/2}}{6 \sqrt{2}} + 
                O(\bar{l}^{\,-3/2})
\nonumber \\ 
        &= & \frac{1}{12} + \frac{T^{\alpha}}{6 \lambda \sqrt{2}} + O(T^{3
  \alpha}), 
\end{eqnarray}
where for the third line $\bar{l}$ was substituted by Eq. (\ref{length}).

\begin{figure}[h]
\begin{center}
\epsfxsize=85mm
\epsffile{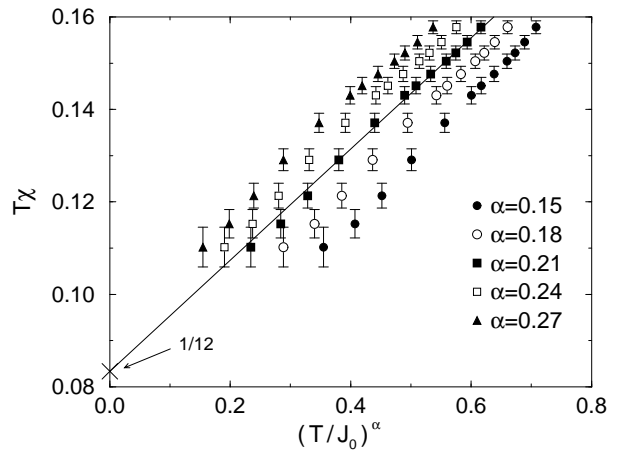}
\end{center}
\caption[]{Uniform Curie constant per spin as a function of
  $T^\alpha$ for different $\alpha$. The best linear behavior
  [Eq. (\ref{curie2})] is found for $\alpha=0.21\pm0.02$. The solid
  line shows a fit of form (\ref{curie2}) for $\alpha=0.21$ and
  fixed intercept $1/12$. This leads to a value of $\lambda=1.0\pm
  0.1$ .} \label{curie_t3}
\end{figure}

Fitting the two leading terms of Eq. (\ref{curie2}) to the QMC data
gives an estimate of the scaling exponent $\alpha$ and the
proportionality factor $\lambda$. Fig.~\ref{curie_t3} shows $T\chi$ as
a function of $T^\alpha$ for different $\alpha$. Using the knowledge
of the exact zero-temperature value $1/12$ of $ T \chi $ we obtain
from this plot $\alpha=0.21 \pm 0.02$, in good agreement with the RSRG
result ($ 0.22 \pm 0.01 $).\cite{west} The proportionality factor is
determined as $\lambda=1.0 \pm 0.1$. These results confirm that our
QMC results for chains of length $L=400$ are not affected by
finite-size effects, because according to Eq. (\ref{length}) the
average length of the correlated clusters, even at the lowest
simulated temperature $J_0 /1000$, is only $\bar{l}\approx 18$, and
therefore much smaller than $L$.

\subsection{Specific heat and entropy}

\begin{figure}[t]
\begin{center}
\epsfxsize=88mm
\epsffile{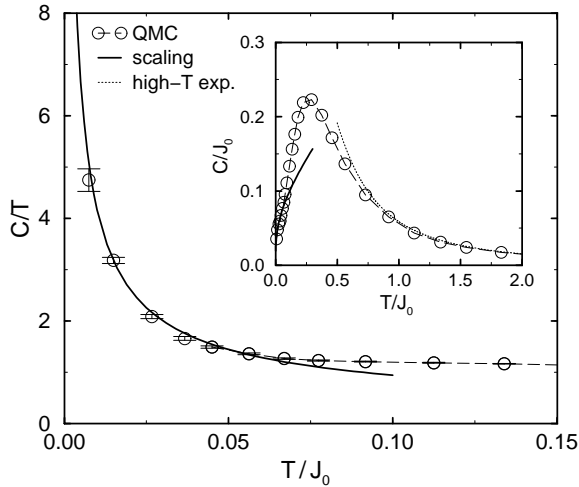}
\end{center}
\caption[]{Specific heat divided by temperature $T$ as a function
  of $T$. The circles are the QMC results, while the solid line was
  calculated as the derivative of Eq. (\ref{entropyR}) with respect to
  $T$. The inset shows the QMC results for the specific heat as a
  function of temperature over a large temperature range. The
  error bars are smaller than the symbols. } \label{specificheat}
\end{figure}

The specific heat $C$ is determined as the numerical derivative of
the internal energy calculated in the QMC simulations.  The errors in
the specific heat increase significantly with lowering temperature.
Nevertheless, $C$ could be determined reliably down to $T\approx
J_0/150$. The QMC results for the specific heat and the ratio $C/T$ as
a function of $T$ are shown in Fig.~\ref{specificheat}.

Since $C/T$ is the derivative of the entropy $\sigma$ with respect to
$T$, we may compare it to a calculation of $\sigma$ by our statistical
cluster analysis. Again we assume that for given temperature each
correlated cluster behaves independently. Hence a cluster of length $
l $ contributes
\begin{equation}\label{entropy1}
\sigma_l=\sum_{S=0}^{l/2} \rho_l(S) \log(2S+1).
\end{equation}
to the entropy.  Here it is necessary to use the discrete expressions
for the distribution functions because large deviations arise in the
continuum approach using Eq. (\ref{dist_cont}), instead of Eq.
(\ref{distpdisc}) and Eq. (\ref{distSdisc}) (see Appendix A).

The entropy per site of the infinite chain is
\begin{equation}\label{entropyR}
\sigma(T)=\frac{1}{\bar{l}}\sum_{l=1}^\infty p_{\bar{l}}(l)\, \sigma_l ,
\end{equation}
where $\bar{l}$ is given by Eq. (\ref{length}).  Using the values of
$\alpha$ and $\lambda$ determined above, the entropy and hence $C/T$
may be calculated from Eq. (\ref{entropyR}). The result is shown as the
solid line in Fig.~\ref{specificheat}. We find good agreement with the
QMC results for the low-temperature range. We emphasize that this
is no fit since all free parameters have been
determined via the uniform susceptibility. At higher
temperatures this approach fails, because the scaling behavior 
Eq.(\ref{length}) of $ \bar{l} $ is no longer valid.

As a further consistency check we estimate the
area below the curve $C/T$ from $T=0$ to $\infty$ using
$\sigma(\infty)=\sigma(T^*)+\int_{T^*}^\infty dT \, (C/T)$, where we
set $T^* =0.03 J_0$. The first term is obtained using Eq.
(\ref{entropyR}) and the integral from $T^*$ to $ \infty $ is
determined numerically, using the QMC results and a high temperature
expansion up to third order. We find $\sigma(\infty)=0.68 \pm 0.01$,
in very good agreement with the expected result $\ln 2$. The fraction
$\sigma(T^*)$ of the entropy is large, contributing approximately 25\%
of the total. Therefore the fact that we find the correct value for
$\sigma(\infty)$ is a further convincing test for the correctness of Eq.
(\ref{entropyR}), and of our statistical treatment of the scaling
regime.

\subsection{Generalized staggered susceptibility}

The generalized staggered susceptibility is defined here as the linear
response of the spin chain to a staggered field $ H_{\rm st} $ whose sign on
site $ j $ is given by 

\begin{equation}
\tau_j = \prod^{i-1}_{m=0} {\rm sgn}(-J_m)
\end{equation}
in analogy to the regular AFM staggering. Based on the discussion of
the non-linear sigma model version of the random spin chain Nagaosa and 
coworkers suggested that each cluster forms a (classical) staggered
spin $ s_{\rm st} $ proportional to the cluster length $ l $,\cite{nagaosa}

\begin{equation}
s_{\rm st} = \zeta l \qquad \qquad \zeta = \mbox{constant}
\label{sst}
\end{equation}
Then the staggered field $ H_{\rm st} $ couples to $ s_{\rm st} $
analogous to the case of a uniform field and the susceptibility should
be given by the following low-temperature scaling form,

\begin{equation}
\chi_{\rm st} \propto \frac{\bar{s}_{\rm st}^2}{T \bar{l}} \propto
\frac{\bar{l}}{T} \propto T^{-(1 + 2 \alpha)}.
\label{cst}
\end{equation}

Using the proposed Eq. (\ref{sst}) and performing a similar analysis as
for the uniform susceptibility, we obtain the following form
\begin{equation}\label{chist}
\chi_{\rm st}=\frac{\zeta^2 \lambda^2}{3} T^{-(1+ 2 \alpha)}+
\frac{\zeta-2\zeta^2}{3} T^{-1}+O(T^{-1+2 \alpha}).
\end{equation}

However, no $\zeta$ can be found such that Eq.
(\ref{chist}) fits our QMC data. Indeed, the low-temperature behavior
of $ \chi_{\rm st} $ is governed by a leading powerlaw $ \chi_{\rm
  st} \propto T^{-\gamma} $ with $ \gamma = 1.17 \pm 0.01 $ which is
clearly different from the anticipated $ 1 + 2 \alpha $. One may argue
that the temperatures covered by the QMC simulations are not low
enough and that only at lower temperature the exponent will approach
the value $1+2 \alpha$. The numerical data, however, show no
indication for such an increase of the exponent down to temperatures
as low as $T=J_0/1000$.

\begin{figure}[h]
\epsfxsize=85mm
\begin{center}
\epsffile{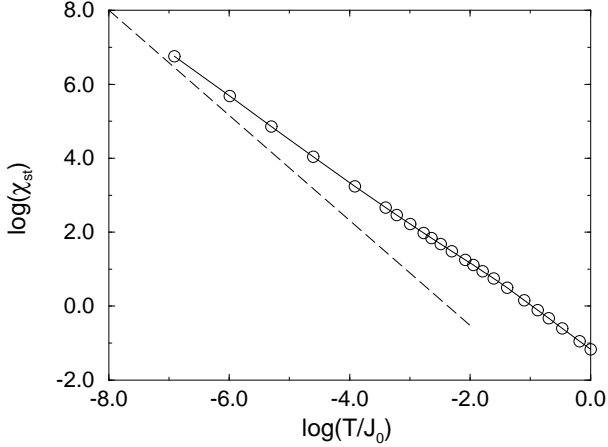}
\end{center}
\caption[]{Double logarithmic plot of the generalized staggered 
  susceptibility as a function of temperature. The error bars are
  smaller than the symbols. The dashed line shows a divergence with
  $T^{-(1+2\alpha)}$.}
\label{fig_stagg_susc}
\end{figure}

This leads to the conclusion, that
the assumption in Eq.~(\ref{sst}) and Eq.~(\ref{cst}) is not
appropriate, and a revision of the ideas on the generalized staggered
susceptibility is necessary. 

In the following we will follow an alternative path which includes
quantum effects ignored above and finally gives an overall consistent
view of this problem. We consider the application of the staggered
field $ H_{\rm st} $ as a perturbation described by addition the
following term to the Hamiltonian,
  
\begin{equation}
\hat{V}_{\rm st} = -H_{\rm st} \hat{S}_{\rm st} 
= -H_{\rm st} \sum_j \tau_j S^z_j 
\end{equation}
We now focus on the contribution of a given correlated cluster of
length $ l $ to the generalized staggered susceptibility. Since the
cluster is correlated it is described by its ground state $ | \Psi_0
(S,M) \rangle $ where $ (S,M) $ are the spin quantum numbers
(effective spin degree of freedom). This state is, in general, not an
eigenstate of $ \hat{V}_{\rm st} $ and we find that $ \langle \Psi_0
(S,M) | \hat{S}_{\rm st} | \Psi_0 (S,M) \rangle $ grows with cluster
size at most as $ l^{1/2} $ (see Appendix B). As we will see below
this leads to a Curie or sub-Curie law of the generalized staggered
susceptibility. Therefore, we include the second order in
perturbation,

\begin{eqnarray}
\Delta E_M & = & - H_{\rm st} \langle \Psi_0 (S,M) |
\hat{S}_{\rm st} | \Psi_0 (S,M) \rangle \nonumber\\
& &\quad \displaystyle\qquad
- H^2_{\rm st} \sum_n \frac{|\langle
\Psi_0 (S,M) | \hat{S}_{\rm st} | \psi_n \rangle |^2}{\epsilon_n -
\epsilon_0 } \\ 
& \equiv & - H_{\rm st} e_{M1} - H^2_{\rm st} e_{M2}, 
\end{eqnarray}
where $ |\psi_n \rangle $ denotes the excited state with energy $
\epsilon_n $ and  $ \epsilon_0 $ is the ground state energy.

The temperature dependence of the generalized staggered susceptibility
is easily calculated from the free energy of the cluster,

\begin{equation}
F_S= - k_B T {\rm ln} Z_S, \qquad Z_S = \sum^S_{M=-S} e^{-\Delta E_M /
k_B T} .
\end{equation}
Note that other contributions to the partition function $Z $ are of
the order $ e^{-\Delta_0 / k_B T} $ and can be ignored ($ \Delta_0 $
($ \sim k_B T $) the excitation gap of the cluster).  The generalized
staggered susceptibility per spin in the cluster is then defined as

\begin{equation}
\chi_{\rm st} = \left. \frac{1}{l} \frac{\partial^2 F}{\partial H^2_{\rm st}}
\right|_{H_{\rm st}=0} = \frac{1}{(2S+1)l} \sum^S_{M=-S}
(\frac{e_{M1}^2}{k_B T} + 2 e_{M2} )
\end{equation}
Now let us give upper bounds to the two terms. For $ e_{M1} $ we find

\begin{equation}
e_{M1} \leq g_1 l^{\rho/2}
\label{estim1}
\end{equation}
independent of $ M $ with a constant $ g_1 $ and the exponent $ \rho
/2 $ (see Appendix B). This upper bound is connected with the
following estimate of $ e_{M2} $ which contains two elements the
denominator and the numerator. The denominator has a simple lower
bound $ \epsilon_n - \epsilon_0 \geq \Delta_0$. An upper bound of the
numerator is obtained if the sum over $n$ is extended to run also over
the states $ |\Psi_0 (S,M)\rangle$, leading to

\begin{equation} \begin{array}{ll} 
e_{M2} & \displaystyle \leq \frac{\langle \Psi_0 (S,M) | \hat{S}_{\rm st} 
\hat{S}_{\rm st} | 
\Psi_0 (S,M) \rangle }{\Delta_0} \\ & \\ & \displaystyle 
= \frac{1}{\Delta_0} \sum_{i,j}
\tau_i \tau_j \langle \Psi_0 (S,M) | S^z_i S^z_j  | \Psi_0 (S,M)
\rangle. \\ & 
\end{array}
\end{equation}
This leads us to the discussion of the generalized staggered
correlation function which we assume to be bounded as

\begin{equation}
\sum_{i,j}
\tau_i \tau_j \langle \Psi_0 (S,M) | S^z_i S^z_j  | \Psi_0 (S,M)
\rangle \leq g_2 l^{\rho} 
\label{estim2}
\end{equation}
where $ g_2 $ is a constant and $ \rho $ a scaling exponent. It is
easy to see that $ \rho=2 $ for ``long-range ordered'' clusters and $
\rho=1 $ for short range correlation. An intermediate $ \rho $
indicates a correlation function which decays with a powerlaw,

\begin{equation}
\Gamma_0 (r) = \sum_i \tau_{i} \tau_{i+r} \langle \Psi_0 (S,M) | S^z_i
S^z_{i+r}  | \Psi_0 (S,M ) \rangle \propto r^{- \eta} 
\label{corr}
\end{equation}
for large $ r $ and $ l $ what leads to $ \rho = 2 - \eta $. Combining
these estimates the generalized staggered susceptibility gets the
following upper bound from the two terms, $ \chi_{\rm st} = \chi_1 +
\chi_2 $,

\begin{equation} \begin{array}{l} \displaystyle 
\chi_1 = \sum_M \frac{e_{M1}^2}{(2S+1) l k_B T} \leq
\mbox{const}\, \frac{l^{\rho-1}}{T} \propto T^{-(1+2\alpha(\rho - 1))},
\\ \\ \displaystyle 
\chi_2 = \sum_M \frac{e_{M2}}{(2S+1)l} \leq
\mbox{const}\, \frac{l^{\rho-1}}{\Delta_0} \propto T^{-(1+2\alpha(\rho -1))},
 \\
\end{array} \end{equation}
where we took $ \Delta_0 \sim k_B T $ and $ l \sim \bar{l} \propto
\Delta^{-2 \alpha}_0 $. Since $ \eta=0 $ 
is a lower bound for $ \eta $ we find that the exponent of $ \chi_{\rm st}
\propto T^{- \gamma} $ lies between 1 and $ 1 + 2 \alpha
$. Furthermore the knowledge of $ \gamma $ from our numerical
calculation allows us now to give an upper bound on the exponent $
\eta $ ($ \gamma = 1.17 \pm 0.01 $ from Fig.~\ref{fig_stagg_susc}),

\begin{equation} \left.
\begin{array}{l} 
\gamma \leq 1 + 2 \alpha (\rho -1) \\ \rho = 2 - \eta \end{array}
\right\} \to \quad \eta \leq 1 -
\frac{\gamma-1}{2 \alpha} \approx 0.62 \pm 0.02
\label{eta_estim}
\end{equation}
This result is consistent with a recent calculation of the correlation
function of large clusters using the density matrix renormalization
group method by Hikihara and coworkers.\cite{hiki} Determining the
correlation function of the ground state averaged over many samples
they found a powerlaw with $ \eta \approx 0.4 - 0.5 $. We will also
see that our following data fit well into this interpretation. Hence
we can conclude that the correlation is longer ranged than in a
regular AFM spin-1/2 chain where the exponent is 1. On the other hand,
the exponent of $ \chi_{\rm st} $, smaller than $ 1 + 2 \alpha $, is
inconsistent with the assumption that the ground state of the random
FM-AFM chain has long range order.

\subsection{Correlations and generalized staggered magnetization}

\begin{figure}[h]
\epsfxsize=85mm
\begin{center}
\epsffile{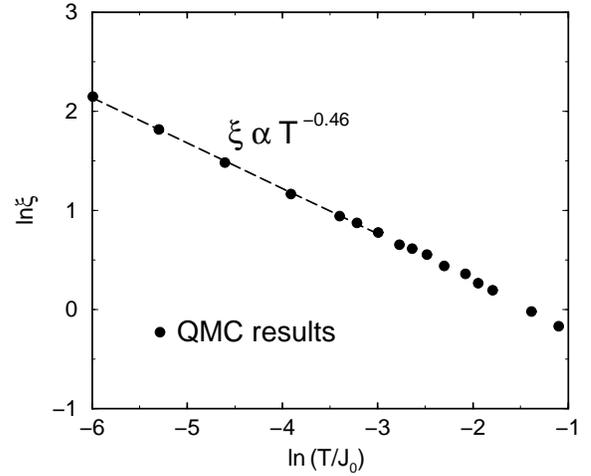}
\end{center}
\caption[]{Double logarithmic plot of the correlation length $\xi$ as 
  function of temperature $T$. $\xi(T)$ diverges approximately with
  $T^{-0.46}$ (dashed line) for $T\rightarrow 0$. The error bars are
  smaller or of order of the symbols.}\label{corrlengthFig}
\end{figure}

Finally, we calculate the correlation function
\begin{equation}
\Gamma(r)= \langle \frac{1}{L}\sum_{i=1}^L S_i^z S_{i+r}^z \left(
\prod_{m=i}^{i+r-1} \mbox{sgn}(-J_m)\right) \rangle
\end{equation}
for temperatures $J_0/400 < T < J_0$, using the QMC algorithm. For
large $r$ and at fixed temperature $T$, the correlation function
$\Gamma(r,T)$ is found to be rather well described by a pure
exponential form
\begin{equation}\label{corr_form}
\Gamma(r,T)= R(T) e^{-r/\xi(T)},\ \ r \gg \xi(T).
\end{equation}
$\xi(T)$ is the correlation length and the prefactor $R(T)$ is $r$
independent. Fitting Eq. (\ref{corr_form}) to the QMC
data, leads to an estimate of $\xi(T)$ and $R(T)$. The 
correlation length as function of temperature is shown in
Fig.~\ref{corrlengthFig} which diverges for $T\rightarrow 0$ with the 
approximate powerlaw,
\begin{equation}\label{corr_exp}
\xi(T)\propto T^{-0.46} .
\end{equation}

Following the concept of our statistical cluster analysis, the
correlation length $\xi(T)$, a measure of distance within spins are
correlated in the random spin chain, should be proportional to the average
cluster length $\bar{l}$. From this point of view we therefore expect
a low temperature behavior according to
\begin{equation}
\xi\propto\bar{l}\propto T^{-2\alpha}.
\end{equation}
Putting in the value $\alpha=0.21\pm0.02$, determined before, a
divergence of $\xi(T)\propto T^{-0.42\pm 0.04}$ is predicted,
in reasonable agreement with the observed divergence, Eq. (\ref{corr_exp}).
After considering the length scale of the correlation function we now
turn to the amplitudes. 

We propose the following scaling behavior for the correlation function
of random FM-AFM spin chains (for $T\rightarrow 0$):
\begin{equation}\label{corr_scal}
\Gamma(r)=\xi^\nu\, \tilde\Gamma(r/\xi),\ \ (\frac{r}{\xi}\gg 1)
\end{equation}
where $\tilde\Gamma(x)$ is a universal (temperature independent)
function.

\begin{figure}[h]
\begin{center}
\epsfxsize=85mm
\epsffile{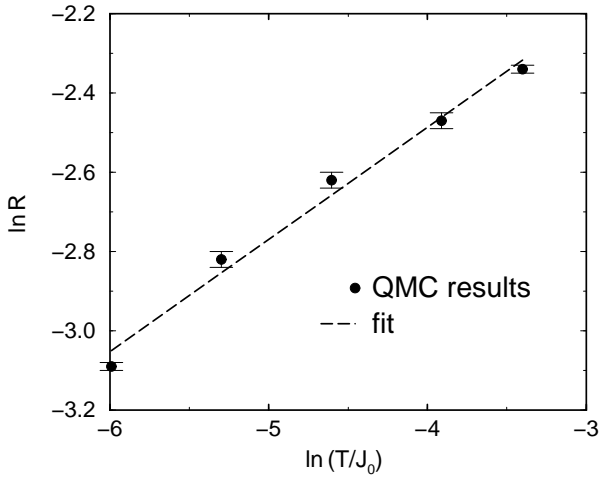}
\end{center}
\caption[]{Double logarithmic plot of the prefactor of the correlation 
function  $R$ as function of temperature $T$ [cf. Eq. (\ref{corr_form}) 
and Eq. (\ref{R})]. The slope is $-2\nu\alpha=0.28\pm 0.02$ (dashed line), 
leading to an estimate of $\nu=-0.61 \pm 0.06$. }\label{log_eta}
\end{figure}

As a consequence, comparing Eq. (\ref{corr_form}) to Eq.
(\ref{corr_scal}), at low temperatures, the prefactor $R(T)$ should
behave as $\xi(T)^\nu$, leading to
\begin{eqnarray}\label{R}
\ln R(T)-\ln R(T_0)&=&\nu\ln\xi-\nu\ln\xi_0 \nonumber \\
                     &=&-2\nu\alpha\ln T+2\nu\alpha\ln T_0,
\end{eqnarray}
where $R(T_0)$ and $\xi_0$ is the prefactor of the correlation and 
the correlation length, respectively, at fixed temperature $T_0$ [c.f.
Eq.  (\ref{corr_form})]. In Fig.~\ref{log_eta} the logarithm of the
prefactor $R$ is shown as function of $\ln T$, at very low
temperatures. The slope $-2\alpha\nu$ is $0.28\pm 0.02$, leading to a
first estimate of $\nu=-0.61 \pm 0.06$.

Another possibility to estimate the exponent $\nu$ is to consider the
low-temperature behavior of the square of the generalized staggered
magnetization defined by
\begin{eqnarray}
M_{\rm st}^2&=& \langle \left( \frac{1}{L}\sum_{i=1}^L
\tau_iS_i^z\right)^2  \rangle \\
            &=&\frac{2}{L}\left(\sum_{r>0}\Gamma(r)
+\frac{\Gamma(0)}{2}\right) \label{Mst1}\\   
            &\approx & \frac{2}{L}\int_0^\infty dr \Gamma(r)\ \ (L \gg
\xi\gg 1), 
\label{Mst}
\end{eqnarray}
where we replaced the sum by an integral and assumed $\xi\gg1$ (i.e.
very low $T$).  This allows us to neglect the term $\Gamma(0)/2L$.

Using the scaling form Eq. (\ref{corr_scal}) for $\Gamma(r)$ in
Eq. (\ref{Mst}) we obtain,
\begin{eqnarray}
M_{\rm st}^2&=&\frac{2}{L}\xi^{\nu}\int_0^\infty dr \tilde\Gamma(r/\xi) \\
            &=&\frac{2}{L}\xi^{\nu+1}\int_0^\infty dx \tilde\Gamma(x).
\end{eqnarray}
As the function $\tilde\Gamma(x)$ is temperature independent, the square 
of the generalized staggered magnetization should diverge as
\begin{equation}
M_{\rm st}^2\propto\xi^{\nu+1}\propto T^{-2 \alpha (\nu+1)}.
\end{equation}

\begin{figure}[h]
\begin{center}
\epsfxsize=85mm
\epsffile{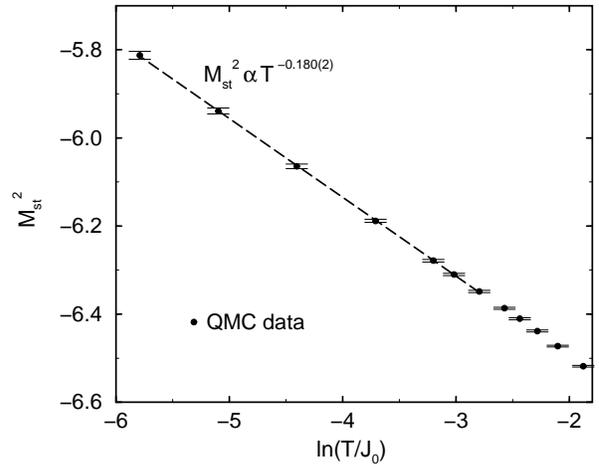}
\end{center}
\caption[]{Double logarithmic plot of the square of the generalized 
staggered magnetization $M_{\rm st}^2$ as function of temperature $T$. 
$M_{\rm st}^2$ diverges with $T^{-0.180\pm 0.002}$ (dashed line) for 
$T\rightarrow 0$.}\label{stagg_magn} 
\end{figure}

The QMC data of the square of the generalized staggered susceptibility
is shown in Fig.~\ref{stagg_magn}. The numerical results show a
divergence of $M_{\rm st}^2\propto T^{-0.180\pm 0.002}$. From this we
find for the exponent $\nu\approx -0.61$ which is in good agreement
with the previous result, deduced from the prefactor of the
correlation. Note, however, that we estimate a rather large error of
about 15\% for $\nu$,as a consequence of the uncontrolled
approximation performed by replacing the sum by an integral in Eq.
(\ref{Mst1}) and neglecting the term $\Gamma(0)/2L$.

Our QMC results show that the scaling behavior of the correlation
function $\Gamma(r)$ of the FM AFM random spin chains is of the form of
Eq. (\ref{corr_scal}). The exponent $\nu$ is roughly estimated
$\nu=-0.61 \pm 0.06$ and is a third exponent (apart from $\alpha$ and
$\gamma$) describing the low temperature scaling regime of random FM-AFM
spin chains.

The exponent $ \nu $ is related to the correlation function used for
the estimate of the generalized staggered susceptibility. We would
like to give here a brief argument for the scaling form in Eq.
(\ref{corr_scal}). We consider a given distribution of clusters for
fixed temperature, and average cluster length $ \bar{l} $ with the
distribution function in Eq. (\ref{dist_cont}). From the assumption
that the clusters are completely uncorrelated we derive the following
approximation for the generalized staggered correlation function for
distance $ r \gg \bar{l} $

\begin{equation}
\Gamma(r) \propto \int^{\infty}_r dl p_{\bar{l}} (l) r^{- \eta} 
\propto r^{- \eta} e^{-r/ \bar{l}}. 
\label{estcor}
\end{equation}
Here only clusters of length larger than $ r $ can contribute and give
on the average a value scaling with $ r $ as $ r^{-\eta} $ [ground
state correlation within the cluster, see Eq. (\ref{corr})].  Naively
we might assume that $ \eta $ and $ \nu $ are identical. However, the
numerical calculations show that the coherence length is shorter than
the average cluster length ($ \xi \approx 2 \bar{l}/3 $) although
their temperature dependence is apparently the same. The form given in
Eq. (\ref{estcor}) may be interpreted as an upper bound which does not
account properly for the thermal fluctuation effects in the (most
important) long clusters whose excitation gap is of the same order as
the temperature.

\section{Conclusions and Summary}

In the study of the random FM-AFM spin-$1/2$ chain, the continuous
time version of the powerful quantum Monte Carlo loop algorithm
allowed us to reach temperatures where clear characteristics of the
universal low-energy properties of this system are observable. The
characteristic temperature where the scaling behavior begins depends
on the initial distribution of bond strengths.  The choice of a
non-singular distribution has the advantage that this temperature
regime is well accessible with our simulation techniques. The key
feature for analyzing the low-temperature data lies in the fact that
the spins are correlated within clusters whose length grows with
decreasing temperature. The scaling of the average length
$\bar{l}=\lambda^2 (T/J_0)^{-2\alpha}$ and the statistics of the
effective spin distribution allow a very good analysis of the
numerical data.  We determine both the prefactor and the exponent. The
latter is assumed to be universal and agrees surprisingly well with
the exponent found by the RSRG, $ \alpha = 0.21 \pm 0.02
$.\cite{west1,west} These two parameters ($\alpha$ and $\lambda$) are
sufficient to describe the uniform susceptibility, entropy and
specific heat in the whole low-temperature scaling regime.

A property not accessible by the RSRG method mentioned above is the
generalized staggered susceptibility. In our simulation we show that a
new exponent $ \gamma $ appears, describing the low-temperature
behavior, $ \chi_{\rm st} \propto T^{-\gamma} $ with a value $ \gamma
\approx 1.17 $.  A further exponent $ \nu $ arises from the scaling
form of the correlation function Eq. (\ref{corr_scal}). We have shown
that $ \nu $ ($ \approx 0.61 $) is connected with the ground state
correlation function which probably exhibits a powerlaw decay over
long distances, $ \Gamma_0 (r) \propto r^{-\eta} < r^{-\nu} $. Both
exponents are assumed to be universal like $ \alpha $, since they are
derived from the properties in the low-temperature scaling regime. At
present, however, we cannot estimate the accuracy of the derived
values. There is also no obvious relation between $ \gamma $, $ \nu $
and $ \alpha $.

Since in the low-temperature regime the system consists of effective
spin degrees of freedom which become more and more classical in their
nature \cite{nagaosa}, there is a clear tendency towards order in the
ground state. The question is whether the remaining quantum effects
(fluctuations) are sufficient to destroy long range order, here.
Therefore, the important result, besides confirming the physical
picture obtained by the RSRG scheme, is the fact that we do not find
any indication that the ground state has long range order. The
generalized staggered susceptibility as well as $ M^2_{\rm st} $ show
too increase too slowly at low temperatures. Nevertheless, the
correlations must be clearly longer ranged than in the uniform AFM
quantum spin-1/2 chain whose ground state staggered correlation
function behaves as $ \Gamma_0 (r) \propto r^{-1} $. This finding is
consistent with a recent density matrix renormalization group study
and an improved version of the real space renormalization group scheme
by Hikihara {\it et al.}.\cite{hiki} Their treatment, however,
suggests that the correlation function might reach its real long-range
behavior only at very long distances which we do not access in our
finite temperature calculation ($ \bar{\ell} < 20 $).

Among the (unfrustrated) random spin chains with complete spin
rotation symmetry there are only two distinct classes, the ones with
the random singlet phase fixed point and the others which belong to
the class studied here.\cite{singular} For the former it was shown by
Fisher that the RSRG scheme describes the fixed point behavior 
exactly.\cite{fisher}
This was not possible so far for the random FM-AFM spin
chain. Our numerical and statistical analysis, however, demonstrates
very convincingly the validity of the universal scaling assumption.
Finally, we would like to emphasize that our statistical fitting
procedure is very suitable and useful to analyze not only numerical,
but also experimental results of this class of random spin systems.

\section*{Acknowledgements}

We would like to thank A. Furusaki, P.A. Lee, N. Nagaosa, T.M. Rice
and E. Westerberg for many fruitful discussions. One of us (B.F.) is
also grateful for financial support from the Swiss Nationalfonds. The
calculations were performed on the Intel Paragon at the ETH Z\"urich.

\section*{Appendix A: Continuous vs. discrete distributions}
\addcontentsline{toc}{section}{Appendix A: Continuous vs. 
discrete distributions}

There is one point to which attention has to be paid when
using the
statistical cluster analysis. For each considered observable it is
necessary to investigate if for its calculation the exact form of
the distributions [Eq. (\ref{distpdisc}) and (\ref{distSdisc})]
have to be taken or if the approximate continuous forms Eq.
(\ref{dist_cont}) and the replacement of the sums by integrals is
also reliable. It is clear that the second case is the
more favorable one, since it gives simple results for the
low-temperature behavior of the corresponding physical observable.
Sums over the discrete distributions, on the other hand, usually can
only be evaluated numerically. Therefore, the continuous forms Eq.
(\ref{dist_cont}) are prefered as long as they do not lead
to a loss of precision. This is shown here for the example of the
uniform Curie constant and the entropy.

\begin{figure}[h]
\begin{center}
\epsfxsize=85mm
\epsffile{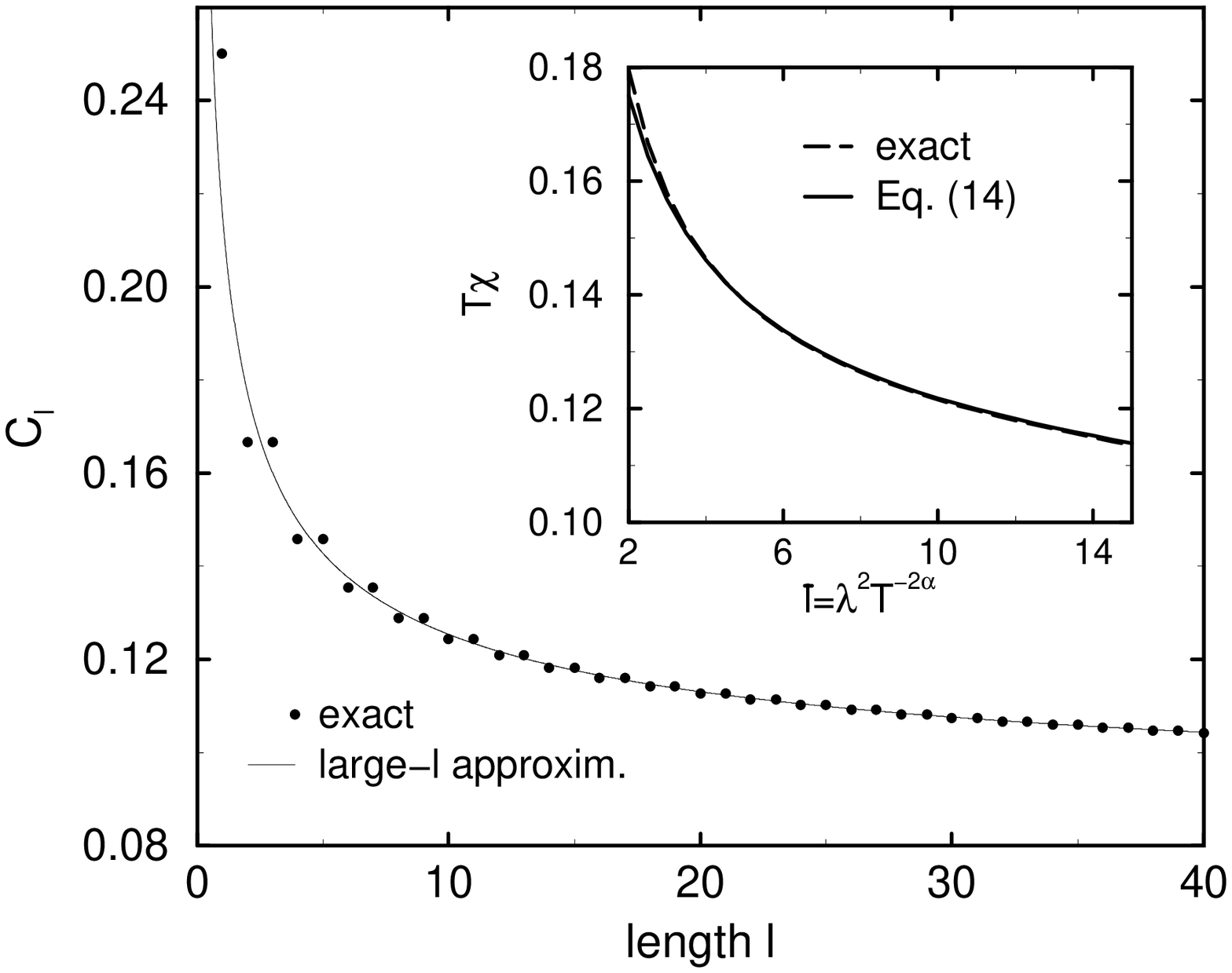}
\end{center}
\caption[]{Uniform Curie constant $\mbox{$\mathcal{C}$}_l$ per site of
  completely correlated clusters of length $l$, calculated exactly
  (solid points) and with large-$l$ approximation (solid line). The
  inset shows the Curie constant of an infinite chain as function of
  the average cluster length $\bar{l}$, calculated exactly (dashed
  line) and with approximate distribution $p_{\bar{l}}$ and
  $\rho_l(S)$ (solid line). For details, see text.}
\label{curie_app}
\end{figure}

We first concentrate on the uniform Curie constant. The Curie constant
for a completely correlated cluster as function of its length is shown
in Fig.~\ref{curie_app}. It is first calculated exactly (solid
points), using Eq. (\ref{curie1}) and the discrete form Eq.
(\ref{distSdisc}) of $\rho_l(S)$. These exact results are compared to
the large-$l$ approximation of $\mbox{$\mathcal{C}$}_l$ (solid line),
which is obtained using also Eq. (\ref{curie1}), but using the
continuous approximation Eq. (\ref{dist_cont}). It can be seen that
this approximation is very good, also for small $l$. As a consequence,
in Eq.  (\ref{curie0}), the approximate large-$l$ value for
$\mbox{$\mathcal{C}$}_l$ can be used without loss of precision.  The
effect of replacing the exact discrete distribution $p_{\bar{l}}(l)$
[Eq. (\ref{distpdisc})] by the continuous one [Eq.  (\ref{dist_cont})]
and the sum by an integral in Eq. (\ref{curie1}), is also
investigated.  In the inset of Fig.~\ref{curie_app} we show the
uniform Curie constant of a random spin chain as a function of the
average cluster length $\bar{l}$. The dashed line represents the
result, obtained by using both exact (discrete) distributions [Eq.
(\ref{distSdisc}) and Eq. (\ref{distpdisc})] and the solid line shows
the result, obtained by using both the continuous distributions Eq.
(\ref{dist_cont}).  We find excellent agreement for $\bar{l}\geq 3$,
what corresponds to temperatures $T\leq J_0/10$. This temperature
range covers the region, where the cluster distribution can be assumed
to be exponential and our statistical cluster analysis results apply.

\begin{figure}[h]
\epsfxsize=85mm
\begin{center}
\epsffile{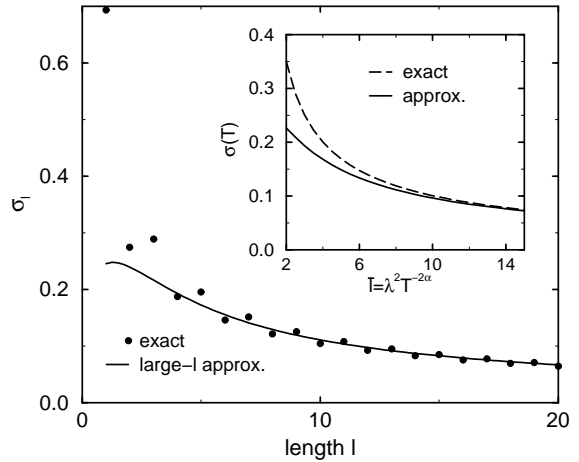}
\end{center}
\caption[]{Entropy contribution $\sigma_l$ of completely correlated
  clusters of length $l$, calculated exactly (solid points) and with
  large-$l$ approximation (solid line). The inset shows the entropy 
  per site of an infinite chain as function of the average cluster
  length $\bar{l}$, calculated exactly (dashed line) and with
  approximate distribution $p_{\bar{l}}$ and $\rho_l(S)$ (solid line).
  For details, see text.}
\label{entropyapp}
\end{figure}

For the entropy, on the other hand, the large-$l$ approximation of
$\sigma_l$ [i.e. using the large-$l$
approximation Eq. (\ref{dist_cont}) for $\rho_l(S)$ instead of the
correct distribution Eq. (\ref{distSdisc}) in Eq. (\ref{entropyR})] 
is very bad for small $l$
(see Fig.~\ref{entropyapp}) and it is not appropriate to use this
approximation for the calculation of $\sigma(\bar{l})$. This can be
seen in the inset of Fig.~\ref{entropyapp}, where we plotted the exact
value $\sigma(\bar{l})$ (dashed line) together with the approximated
value of
$\sigma(\bar{l})$, calculated using the large-$l$ approximation for
$\sigma_l$ (dashed line). The $\sigma(\bar{l})$ based on the
continuum approximation agrees
with the exact one only for $\bar{l}\geq15$, which corresponds to
temperatures $T\leq J_0/500$. In order to compare the statistical
cluster analysis results with the QMC data, it is therefore necessary
to calculate the entropy exactly and not to use the large-$l$
approximation.

\section*{Appendix B: Bounds for the staggered moments}
\addcontentsline{toc}{section}{Appendix B: Bounds for the 
staggered moments}

The aim of this section is to establish the upper bounds for $ e_{M1}
$ and $ e_{M2} $ in Eq. (\ref{estim1}) and (\ref{estim2}),
respectively, for a given cluster with the ground state $ | \Psi_0
(S,M) \rangle $ of degeneracy $ 2 S +1 $. The generalized staggered
correlation function $ \Gamma(r) $ given in Eq. (\ref{corr}) is assumed
to scale with the powerlaw $ r^{-\eta} $. Therefore we find
\begin{equation} \begin{array}{l}
\langle \Psi_0 (S,M) | \hat{S}_{\rm st} \hat{S}_{\rm st} | \Psi_0 (S,M)
\rangle  \\ \\
\qquad = \sum_{i,j} \tau_i \tau_j \langle \Psi_0 (S,M) | S^z_i S^z_j
| \Psi_0 (S,M) \rangle \\ \\ \qquad = 
\int^l_0 dr \int^l_0 dr' \Gamma_{0M}(|r-r'|) \propto l^{2 - \eta} = 
l^{\rho}, \\ 
\end{array} \label{prp} 
\end{equation}
which defines the exponent $ \rho $. The proportionality factor $ g_2
$ in Eq. (\ref{estim2}) was chosen large enough such that $ g_2
l^{\rho} $ acts as an upper bound for all spin quantum numbers $ M $
for given $ S $ and $ l $. 

Now we turn to $ e_{M1} $ for which we obtain an upper bound using
Eq. (\ref{prp}). 

\begin{equation} \begin{array}{ll}
e^2_{M1} & = |\langle \Psi_0 (S,M) | \hat{S}_{\rm st} | \Psi_0 (S,M)
\rangle|^2 \\ & \\ & \leq \sum_{n} \langle \Psi_0 (S,M) | \hat{S}_{\rm st} 
| \Psi_n
\rangle \langle \Psi_n | \hat{S}_{\rm st} | \Psi_0 (S,M)
\rangle \\ & \\ &
= \langle \Psi_0 (S,M) | \hat{S}_{\rm st} \hat{S}_{\rm st} | \Psi_0 (S,M)
\rangle \leq g_2 l^{\rho} \\ &
\end{array} \end{equation}
where the sum runs over the complete set of basis states $ |\Psi_n
\rangle $ including, of course, $ |\Psi_0 (S,M) \rangle $. Taking the
square root we obtain the upper bound for $ e_{M1} $ with $ g_1 \geq
\sqrt{g_2} $.


\begin{thebibliography}{4}
\bibitem{singular} Random spin
  chains whose bond distribution functions are very singular in the
  limit $ J \to 0 $ are excluded from this classification. For these
  systems, usually no universal scaling 
  regime can be reached and the low-temperature behavior is governed
  by the $ J \to 0 $-limit of the initial bond distribution.
\end{thebibliography}
\end{document}